\documentclass{article}
\usepackage[preprint]{spconf}

\usepackage{amsmath,graphicx}
\usepackage{balance}
\usepackage{amsfonts}
\usepackage{bm}
\usepackage{tikz}
       \usepackage{pgfplotstable}
   \usepackage{pgfplots}%  
	\usepgfplotslibrary{units} 
    \pgfplotsset{compat=1.9}
    \makeatletter
    \pgfplotsset{
     unit code/.code 2 args=
       \begingroup
       \protected@edef\x{\endgroup\si{#2}}\x
    }
   \usepackage{siunitx}

\def\x{{\mathbf x}}

\definecolor{cnn_color}{rgb}{0.937,0.161,0.161}
\definecolor{gcc_color}{rgb}{0, 0.4470, 0.7410}
\definecolor{svd_color}{rgb}{0.8500, 0.3250, 0.0980}
\definecolor{wsvd_color}{rgb}{0.9290, 0.6940, 0.1250}

\copyrightnotice{\begin{minipage}{\textwidth}
  \centering
  \tiny
  \copyright~2020 IEEE. Personal use of this material is permitted.  Permission from IEEE must be obtained for all other uses, in any current or future media, including reprinting/republishing this material for advertising or promotional purposes, creating new collective works, for resale or redistribution to servers or lists, or reuse of any copyrighted component of this work in other works.
  \end{minipage}} 

\title{Time Difference of Arrival Estimation from Frequency-Sliding Generalized Cross-Correlations Using Convolutional Neural Networks}

\name{Luca Comanducci$^{\star}$ \qquad Maximo Cobos$^{\dagger}$ \qquad Fabio Antonacci$^{\star}$ \qquad Augusto Sarti$^{\star}$
\thanks{\footnotesize This work has been partially supported by FEDER and the Spanish Government under Grants RTI2018-097045-B-C21
and PRX19/00075.}
}
\address{\small $^{\star}$ Dipartimento di Elettronica, Informazione e Bioingegneria - Politecnico di Milano, via Ponzio 34/5 - 20133 Milano, Italia \\
\small $^{\dagger}$ Departament d'Inform\`atica, Universitat de Val\`{e}ncia, 46100 Burjassot, Spain}
\usepackage{atbegshi,picture}
\AtBeginShipout{\AtBeginShipoutUpperLeft{%
  \put(\dimexpr\paperwidth-1cm\relax,-1.5cm){\makebox[0pt][r]{Paper accepted for presentation in ICASSP 2020}}%
}}

\begin{document}

\ninept

\maketitle

\begin{abstract}

The interest in deep learning methods for solving traditional signal processing tasks has been steadily growing in the last years. Time delay estimation (TDE) in adverse scenarios is a challenging problem, where classical approaches based on generalized cross-correlations (GCCs) have been widely used for decades. Recently, the frequency-sliding GCC (FS-GCC) was proposed as a novel technique for TDE based on a sub-band analysis of the cross-power spectrum phase, providing a structured two-dimensional representation of the time delay information contained across different frequency bands. Inspired by deep-learning-based image denoising solutions, we propose in this paper the use of convolutional neural networks (CNNs) to learn the time-delay patterns contained in FS-GCCs extracted in adverse acoustic conditions. Our experiments confirm that the proposed approach provides excellent TDE performance while being able to generalize to different room and sensor setups.

\end{abstract}
\section{Introduction}
\label{sec:intro}

The estimation of the Time Difference of Arrival (TDoA) between the signal acquired by two microphones is relevant for many applications dealing with the localization, tracking and identification of acoustic sources. The Generalized Cross-Correlation (GCC) with Phase Transform (PHAT)~\cite{knapp1976generalized} has been widely used for this problem and is regarded as a robust method for TDoA estimation in noisy and reverberant environments. Nonetheless, several problems arise in such scenarios, leading to TDoA errors derived from spurious peaks in the GCC due to reflective paths, excessive noise for some time instants, or other unexpected interferers. To mitigate these problems, the authors recently proposed the Frequency-Sliding GCC (FS-GCC), an improved GCC-based method for robust TDoA estimation \cite{Cobos2019FSGCC}. The FS-GCC is based on the analysis of the cross-power spectrum phase in a sliding window fashion, resulting in a set of sub-band GCCs that capture the time delay information contained in different frequency bands. As a result of such analysis, a complex matrix constructed by stacking all the sub-band GCCs is obtained, which can be later processed to obtain a reliable GCC representation, for example, by means of rank-one approximations derived from Singular Value Decomposition (SVD). This paper proposes an alternative processing scheme for FS-GCC representations based on Deep Neural networks (DNNs). 

In recent years, several approaches have been proposed for acoustic source localization using DNNs. Most published methods focus on estimating the Direction of Arrival (DoA) of one or multiple sources considering different kinds of DNN inputs, including magnitude~\cite{NelsonYalta2017} and phase~\cite{chakrabarty2017broadband,adavanne2018direction} information from Short-Time Fourier Transform (STFT) coefficients, beamforming related features~\cite{Salvati2018ExploitingCF}, MUSIC eigenvectors~\cite{takeda2016discriminative,takeda2016sound} and GCC-based features~\cite{xiao2015learning,he2018deep}. End-to-end localization approaches accepting raw audio inputs have also been proposed, using binaural signals for DOA estimation~\cite{vecchiotti2019end} or multi-channel microphone signals for estimating the three-dimensional source coordinates~\cite{vera2018towards}. A fewer amount of works address directly the problem of TDoA estimation. In~\cite{houegnigan}, a multilayer perceptron using the raw signals from a pair of sensors was proposed. More recently, the use of recurrent neural networks to exploit the temporal structure and time-frequency masking was proposed in~\cite{pertila2019time}, accepting log-mel spectrograms and GCC-PHAT features as input.

This work combines the structured information extracted by the FS-GCC with Convolutional Neural Networks (CNNs) to address the TDoA estimation problem in adverse acoustic scenarios. The network is designed as a convolutional denoising autoencoder that acts both as a denoising and as a dereverberation system. The network is trained using two-dimensional inputs corresponding to the magnitude of the FS-GCC matrices extracted from simulated data in noisy and reverberant environments, while the target outputs are defined as the equivalent matrices under ideal anechoic conditions.
We demonstrate the effectiveness of our method by showing performance results in two different rooms over a range of reverberation conditions.
The paper is structured as follows. Section~\ref{sec:signal_model_and_background} describes the signal model and the GCC and FS-GCC approaches. 
Section~\ref{sec:proposed_method} presents the proposed method, while Section~\ref{sec:results} describes the results. Finally, Section~\ref{sec:conclusions} draws some conclusions.

%Pertilä and Parviainen - 2019 - Time Difference of Arrival Estimation of Speech Si (002)

%Vecchiotti et al - End-to-end binaural sound localisation from the raw waveform

%Xiao et al - A learning-based approach to direction of arrival estimation in noisy and reverberant environments

%Vera-Diaz et al - Towards End-to-End Acoustic Localization using Deep Learning

%Takeda et al - Discriminative multiple sound source localization based on DNNs using independent location model

%Takeda - Sound source localization based on DNN with directional activate function exploiting phase information

%Houegnigan - Neural networks for high performance time-delay estimation and acoustic source localization

%He at al - Deep Neural Networks for Multiple Speaker Detection and Localization

%Chackrabarty et al - Broadband DOA estimation using CNNs trained with noise signals

%Adavanne et al - Direction of Arrival Estimation from Multiple Sources using Convolutional Recurrent Neural Networks

%Yalta,  Sound source localizationusing deep learning models

%\begin{itemize}
%    \item Fs-GCC cite archiv paper?
%    \item GCC \cite{knapp1976generalized}
%    \item tdoa deep learning most similar work  \cite{pertila2019time} 
%    \item dereverberation gannot net architecture \cite{ernst2018speech}
%\end{itemize}

\section{Signal Model and Background}
\label{sec:signal_model_and_background}
\begin{figure*}[htb]
\begin{minipage}[t]{.32\linewidth}
  \centering
  \centerline{\includegraphics[width=6.0cm]{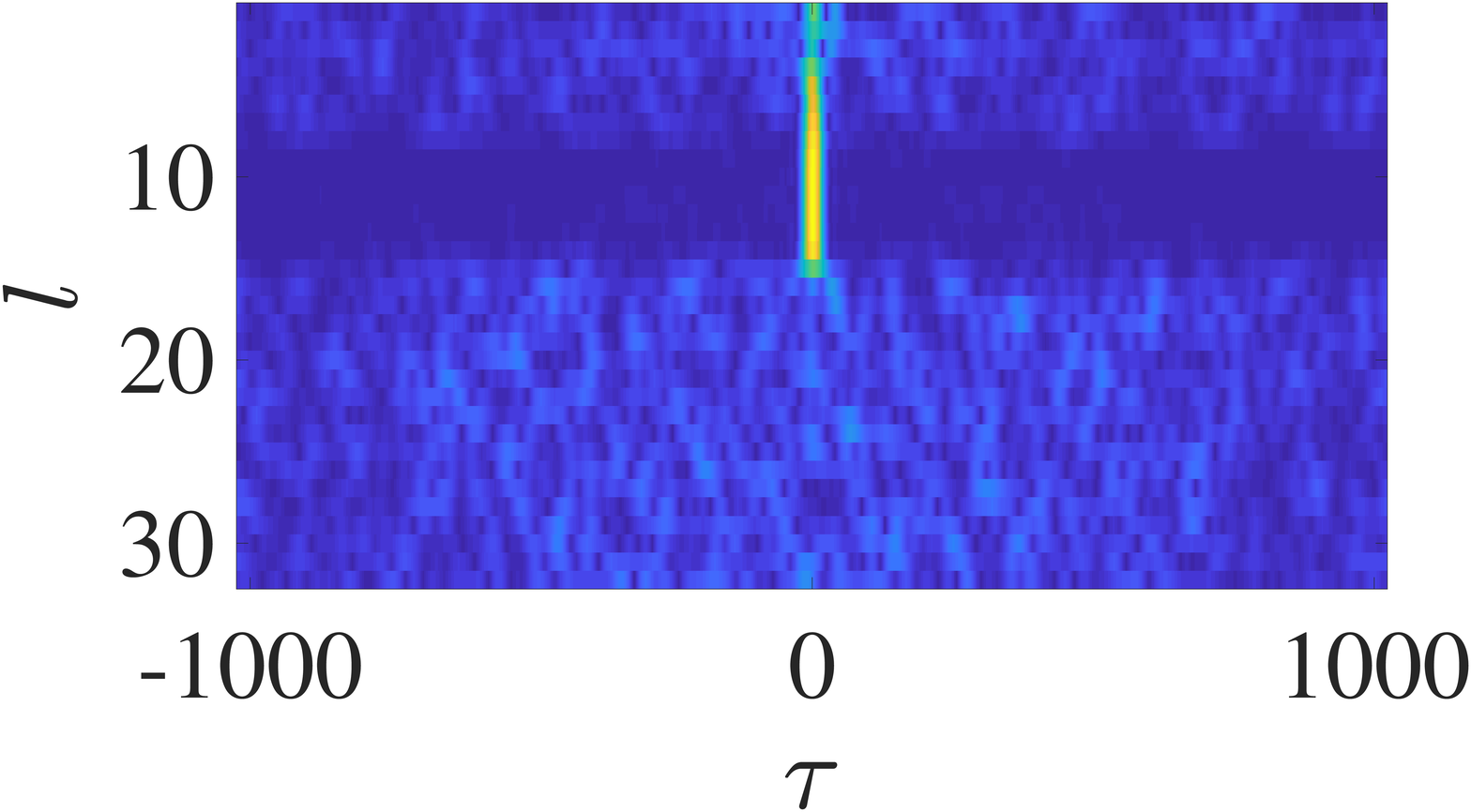}}
  %\vspace{1.5cm}
  \centerline{(a) $\mathrm{T60}=0 ~\mathrm{s}, \mathrm{SNR}=30 ~\mathrm{dB}$}\medskip
\end{minipage}
\begin{minipage}[t]{.32\linewidth}
  \centering
  \centerline{\includegraphics[width=6.0cm]{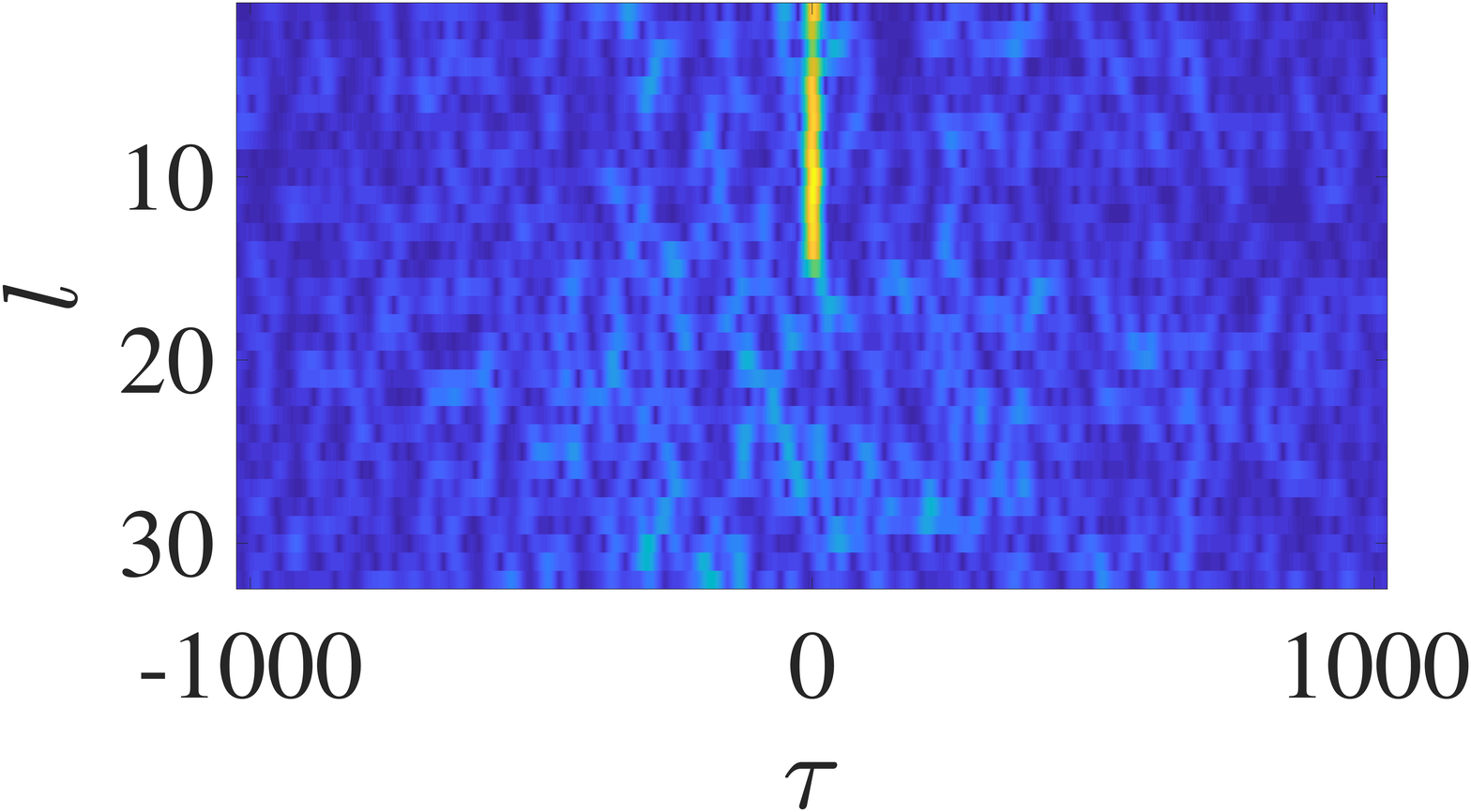}}
 %\vspace{1.5cm}
  \centerline{(b) $\mathrm{T60}=0.2~\mathrm{s}, \mathrm{SNR}=30~\mathrm{dB}$}\medskip
\end{minipage}
\begin{minipage}[t]{.32\linewidth}
  \centering
  \centerline{\includegraphics[width=6.0cm]{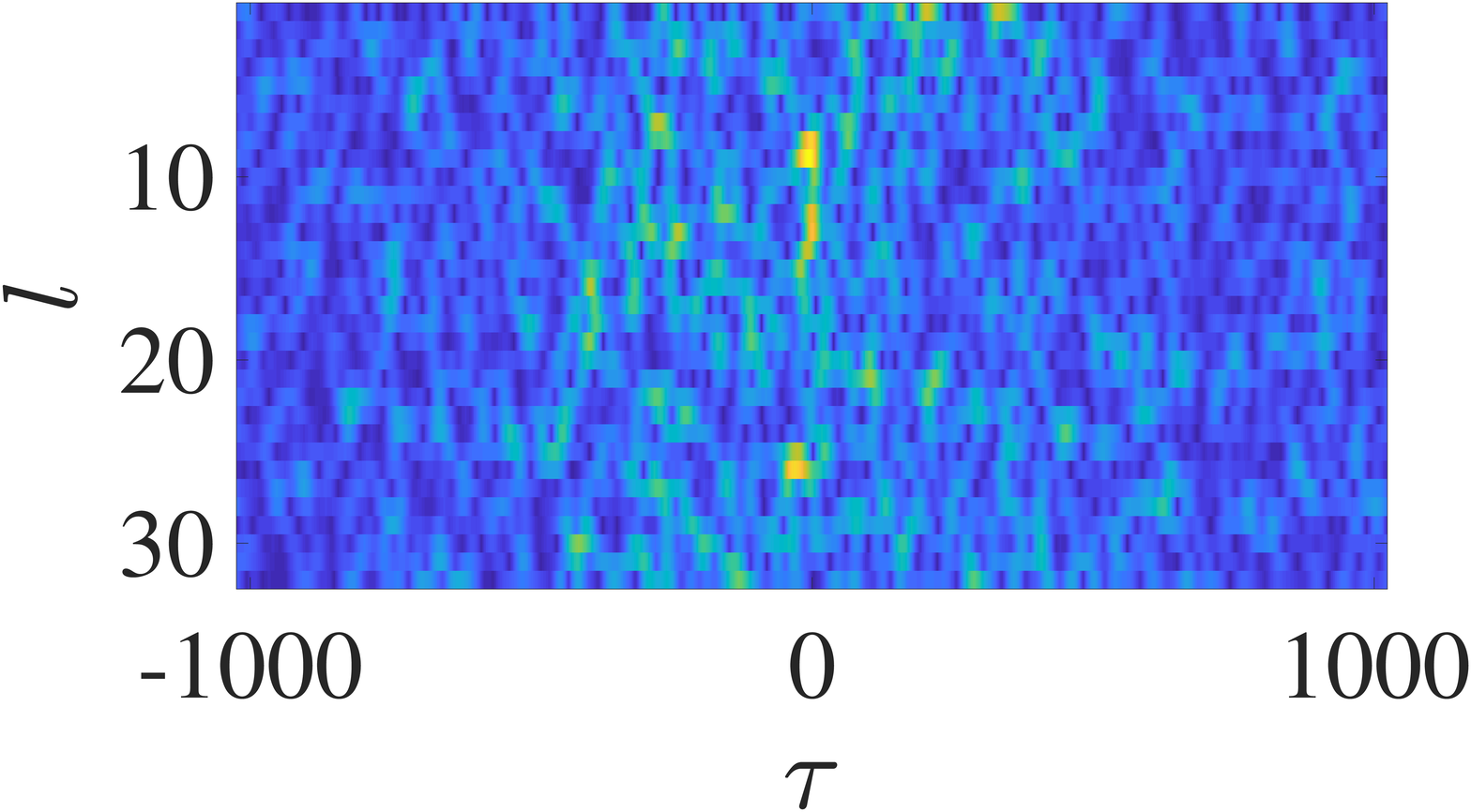}}
  %\vspace{1.5cm}
  \centerline{(c) $\mathrm{T60}=0.5~\mathrm{s}, \mathrm{SNR}=20~\mathrm{dB}$}\medskip
\end{minipage}
\caption{FS-GCC magnitude, $|\mathbf{R}_{12}|$, relative to the same speech signal frame, computed under different conditions of noise and reverberation.}
\label{fig:fsgcc_speech_example}
\end{figure*}

Let us consider a pair of microphones $\mathbf{m}_1,\mathbf{m}_2\in \mathbb{R}^3$ and a source placed in $\mathbf{s}\in \mathbb{R}^3$, emitting a signal $s[n]$, where $n$ is the sample index. The signal received by the two sensors can be described as
\begin{equation}
    \label{eq:mic_signal_time}
    x_m[n] = h_ m[n] * s[n] + w_m[n],\quad m=1,2,
\end{equation}
where $h_m[n]$ is the Room Impulse Response (RIR) response between source $\mathbf{s}$ and microphone $\mathbf{m}_m$ and $w_m[n]$ is an additive noise term.
We can rewrite \eqref{eq:mic_signal_time} in the Discrete-Time Fourier Transform (DTFT) domain as
\begin{equation}
    \label{eq:mic_signal_DTFT}
    X_m(\omega) = H_ m(\omega) \cdot S(\omega) + W_m(\omega),\quad m=1,2,
\end{equation}
where $H_ m(\omega), S(\omega), W_m(\omega)$ are the DTFTs of the RIR, source signal and additive noise signal, respectively.

The TDoA measured in samples and relative to the signal received by microphones $\mathbf{m}_1,\mathbf{m}_2$ and emitted by source $\mathbf{s}$, is defined as 
\begin{equation}
    \label{eq:tdoa}
    \tau_{12} \overset{\Delta}{=} \left\lfloor \frac{||\mathbf{s}-\mathbf{m}_1||-||\mathbf{s}-\mathbf{m}_2||}{c} f_s \right\rceil,% = \eta_1 - \eta_2,
\end{equation}
where $c$ is the speed of sound, $\lfloor \cdot \rceil$ denotes the rounding operator and $f_s$ is the sampling frequency. 
%The values $\eta_1$ and $\eta_2$ represent the time of flight (TOF) in samples of the sound to the sensors. 

\subsection{Generalized Cross-Correlation}
Let us define the phase transform cross-power spectrum $\Psi_{12}(\omega)~\in~\mathbb{C}$ between signals $X_1(\omega)$ and  $X_2(\omega)$ as
\begin{equation}
    \label{eq:pseudospectrum}
    \Psi_{12}(\omega) \overset{\Delta}{=} \frac{X_1(\omega)X_2^\ast(\omega)}{|X_1(\omega)X_2(\omega)|},
\end{equation}
where $^{\ast}$ denotes complex conjugation. Then, the GCC-PHAT (GCC in the following) can be defined as
\begin{equation}
    \label{eq:GCC}
    R_{12}[\tau] \overset{\Delta}{=} \frac{1}{2\pi} \int_{-\pi}^{\pi} \Psi_{12}(\omega)e^{j\omega\tau}d\omega,
\end{equation}
where $\tau$ represents the time delay.
In the ideal anechoic and noiseless case, the GCC will show a unit impulse located at the corresponding true TDOA sample index. Therefore, the time-delay estimation can be performed as
\begin{equation}
    \label{eq:tdoa_est_gcc}
    \hat{\tau}_{12} = \arg \max_\tau R_{12}[\tau].
\end{equation}
When noise and reverberations are taken into account, the GCC presents a series of spurious peaks, which make the time-delay estimation procedure more complex.

\subsection{Frequency-Sliding Generalized Cross-Correlation }
We summarize the frequency-sliding GCC (FS-GCC) method, which improves the GCC method and enables us to analyze how different frequency bands contribute to a direct-path delay estimation. The sub-band GCC relative to an arbitrary frequency band $l$ is defined as 
\begin{equation}
    \label{eq:sub_band_gcc}
    R_{12}[l,\tau] \overset{\Delta}{=} \frac{1}{2\pi} \int_{-\pi}^{\pi} \Psi_{12}(\omega + \omega_l) \Phi(\omega)e^{j \omega \tau} d\omega,
    %\\ 
    %\mathcal{F}^{-1}\{\Psi_{12}(\omega + \omega_l) \phi(\omega)\}
\end{equation}
where $\omega_l$ is the frequency offset corresponding to band $l$. The symbol $\Phi(\omega) \in \mathbb{R}$ corresponds to a symmetric frequency-domain window, centered at $\omega=0$ with frequency support $B_{\phi} \in [0,\pi]$.
A frequency-sliding sub-band GCC can then be obtained by sweeping $\angle{\Psi_{12}(\omega)}$ over possibly overlapping frequency bands: $\omega_l~=~lM_\phi$, $l = 0,\ldots,L-1$, where $M_\phi$ is the frequency hop. The number of bands $L$ can be chosen to cover up for frequencies up to the Nyquist limit: $L = \lfloor (\pi-B_{\bm{\phi}}+M_\phi)/M_\phi\rfloor$.

Let us consider the Discrete Fourier Transform (DFT) $\mathbf{X}_m \in \mathbb{C}^N,~ m=1,2,$ of the microphone signals $x_m[n]$, composed of the coefficients $X_m[k]$, corresponding to the discrete frequencies $\omega_k~=~k \frac{2\pi}{N}$, where $N$ is the length of the DFT. Let us also consider, similarly, the vector $\bm{\Phi} \in \mathbb{R}^N$ containing the samples $\Phi[k]~=~\Phi(\omega_k)$ of the chosen spectral window, symmetrically padded with zeros, in order to contain only $B = \lfloor 2B_{\bm{\phi}} \frac{N}{2\pi}\rfloor$ non-zero elements.

Then, the elements $r_l[n]$ of the sub-band GCC vectors $\mathbf{r}_l \in \mathbb{C}^N$ are obtained by taking the inverse DFT of the windowed PHAT spectrum:
\begin{equation}
    \label{eq: sub-band_GCC_DFT}
    r_l[n] = \frac{1}{N} \sum_{k=0}^{N-1} \frac{X_1^*[k+lM][k+lM]}{|X_1^*[k+lM][k+lM]|} \Phi[k]e^{j \frac{2\pi}{N}kn},
\end{equation}
where $M = \lfloor  M_{\Phi} \frac{N}{2\pi}\rfloor$ is the discrete frequency hop. 

The frequency-sliding GCC (FS-GCC) matrix is constructed by stacking together all the sub-band GCC vectors
\begin{equation}
    \label{eq:fs-gcc}
    \mathbf{R}_{12} = [\mathbf{r}_0^{T}, \mathbf{r}_1^{T}, \ldots, \mathbf{r}_{L-1}^{T}]^{T} \in \mathbb{C}^{L \times N}. 
\end{equation}
In the ideal noiseless and anechoic case, the maximum value of each row of $\mathbf{R}_{12}[l,\tau],~ l = 1,\ldots,L$ is located at the true TDoA $\tau_{12}$. 
When noise and reverberation are present, these will also affect the rows of the FS-GCC, making TDoA estimation harder. A solution, which takes advantage of the FS-GCC properties,is to compute the Singular Value Decomposition (SVD) of $\mathbf{R}_{12}[l,\tau]$ and a low-rank (SVD FS-GCC) or weighted low-rank (WSVD FS-GCC) analyses to estimate the TDOA %to obtain a SVD-based low-rank (SVD FS-GCC) and weighted low-rank (WSVD FS-GCC) approximations of the FS-GCC, explained in detail in
~\cite{Cobos2019FSGCC}. Through these approximations, it is possible to exploit the sub-band representation of the GCC and separate the noise components from the ones related to the direct path.
An example of FS-GCC corresponding to a real speech signal frame, under different amounts of noise and reverberation is shown in Fig.~\ref{fig:fsgcc_speech_example}.
\section{Proposed Method}
\label{sec:proposed_method}
\begin{figure*}[htb]
\begin{minipage}[t]{.32\linewidth}
  \centering
  \centerline{\includegraphics[width=6.0cm]{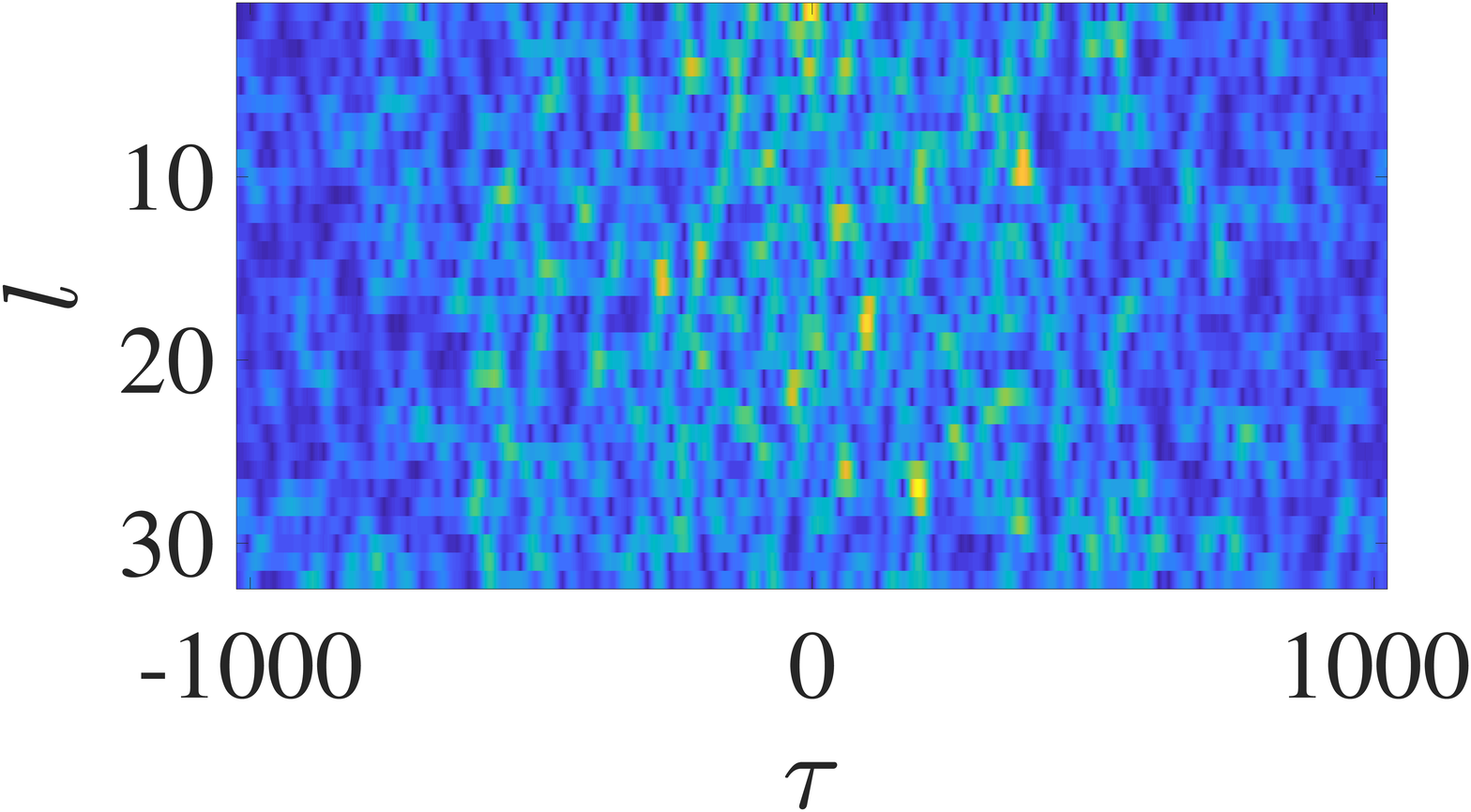}}
  %\vspace{1.5cm}
  \centerline{(a) $|\tilde{\mathbf{R}}_{ij}|$}\medskip
\end{minipage}
\begin{minipage}[t]{.32\linewidth}
  \centering
  \centerline{\includegraphics[width=6.0cm]{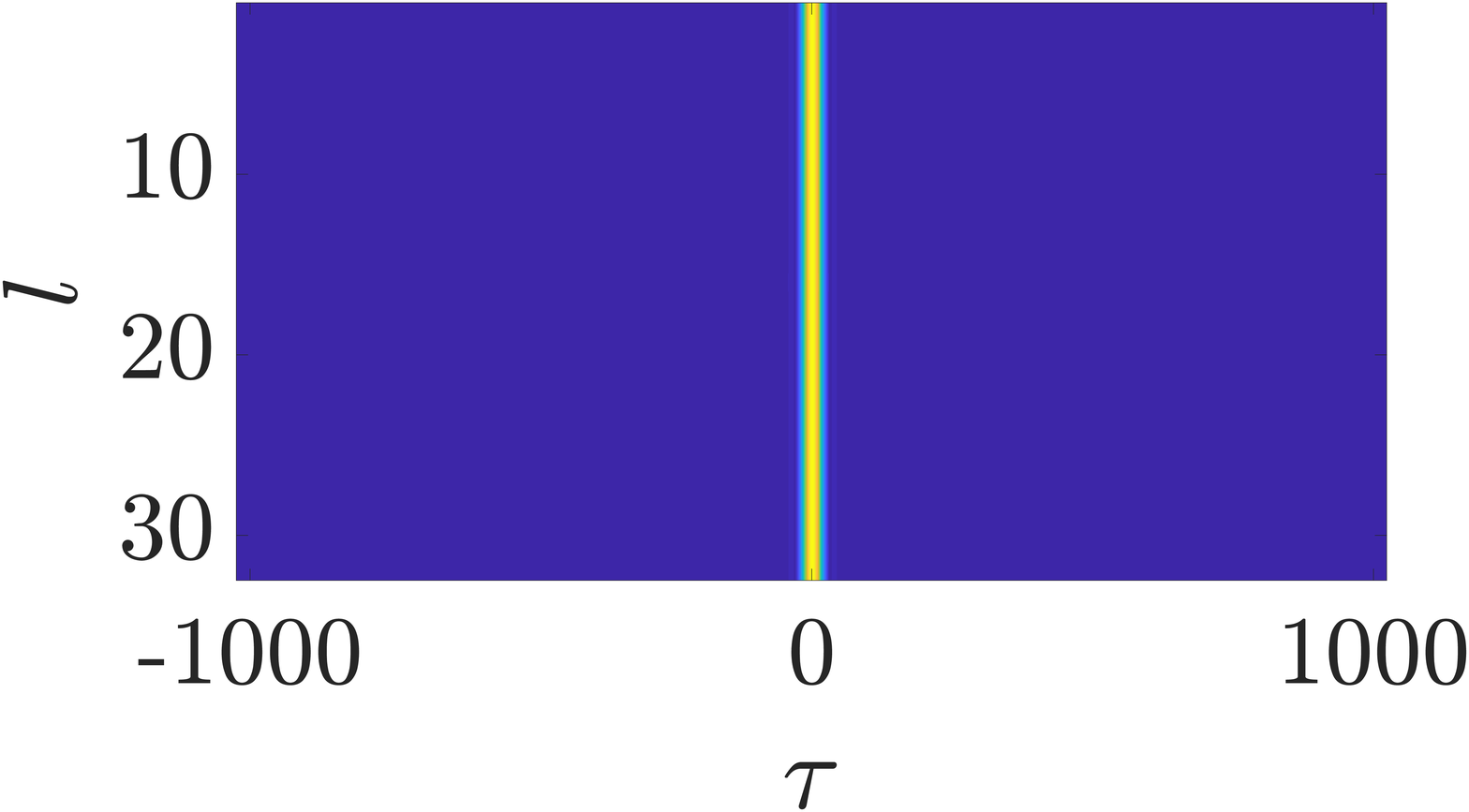}}
  %\vspace{1.5cm}
  \centerline{(b) $|\mathbf{R}_{ij}|$}\medskip
\end{minipage}
\begin{minipage}[t]{.32\linewidth}
  \centering
  \centerline{\includegraphics[width=6.0cm]{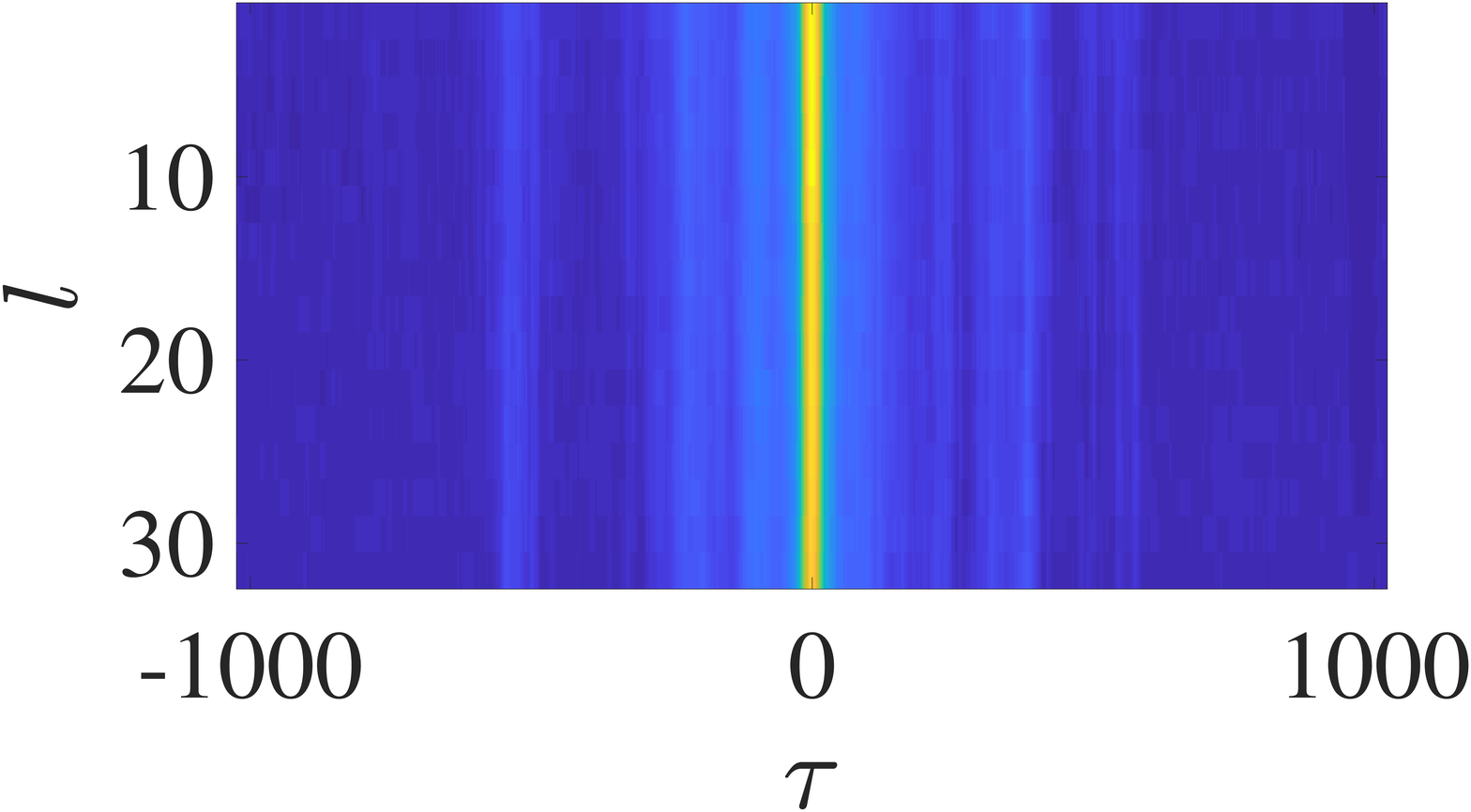}}
 %\vspace{1.5cm}
  \centerline{(c) $|\hat{\mathbf{R}}_{ij}|$}\medskip
\end{minipage}
\caption{FS-GCC (a) computed in a reverberant and noisy environment ($\mathrm{T60} = 1~\mathrm{s}$, $\mathrm{SNR} = 10~\mathrm{dB}$) (b) computed in anechoic and noiseless conditions, (c) reconstructed by our model.}
\label{fig:fsgcc_example}
\end{figure*}

In this section, we describe the proposed model, which consists of a U-Net~\cite{ronneberger2015u} fully convolutional denoising autoencoder. The input is the magnitude of the FS-GCC obtained in a reverberant and noisy environment. The desired output is the magnitude of the corresponding FS-GCC obtained in anechoic and noiseless conditions. %The network thus acts at the same time both as a FS-GCC denoiser and dereverberator .

In the following, we will denote as $|{\mathbf{R}}_{ij}|$, $|\tilde{\mathbf{R}}_{ij}|$ and $|\hat{\mathbf{R}}_{ij}|$ the magnitude of the FS-GCC matrices corresponding to microphones $i$ and $j$, $i \neq j$, obtained in an anechoic environment, in a reverberant one and estimated by the network, respectively. Our objective is, given $|\tilde{\mathbf{R}}_{ij}|$, to obtain an estimate $|\hat{\mathbf{R}}_{ij}|$ of $|\mathbf{R}_{ij}|$ and, thus, to retrieve the matrix function performing the following mapping
\begin{equation}
    \label{eq:obj_function}
    |{\mathbf{R}}_{ij}| =\mathcal{U}(|\tilde{\mathbf{R}}_{ij}|).
\end{equation}
An example of $|\tilde{\mathbf{R}}_{ij}|$, $|\mathbf{R}_{ij}|$ and $|\hat{\mathbf{R}}_{ij}|$, computed for the same source and microphone setup is shown in Fig.~\ref{fig:fsgcc_example}(a), Fig.~\ref{fig:fsgcc_example}(b) and Fig.~\ref{fig:fsgcc_example}(c), respectively. The TDoA estimate $\hat{\tau}_{12}$ is obtained as the time lag corresponding to the maximum of the frequency-band-wise average of the reconstructed FS-GCC, i.e. $\hat{\tau}_{12} = \arg\max_{\tau} \frac{1}{L}\sum_{l=0}^{L-1}\hat{\mathbf{r}}_l^{T}$, where $\hat{\mathbf{r}}_l^{T}$ are the rows of $|\hat{\mathbf{R}}_{ij}|$.

\subsection{Network Architecture}
The proposed network takes as input the $L \times N$ matrix $|\tilde{\mathbf{R}}_{ij}|$ and outputs a matrix of the same size $|\hat{\mathbf{R}}_{ij}|$. $L$ and $N$ can be chosen independently of the architecture, as long as both are a power of $2$.

The autoencoder architecture can be split in two parts: encoder and decoder. The encoder is composed of 4 convolutional layers having the following number of filters: i) $8$, ii) $16$, iii) $32$, iv) $64$. The decoder gets as input the output of the encoder and has a reverse architecture, having $4$ layers with the following number of filters: v) $64$ vi) $32$ vii) $16$ viii) $8$ ix) $1$. Each convolutional layer has stride $(2\times2)$, an asymmetric kernel size of $(10\times5)$ and is followed by Batch Normalization and ReLU. An exception is layer ix) of the decoder, where no Batch Normalization is applied and stride and kernel sizes correspond to ($1 \times\ 1$), in order to reconstruct an image of the same size of the input. Every convolutional layer of the decoder is preceded by an upsampling layer with a factor of $(2 \times 2)$. This choice was made in order to avoid checkerboard artifacts produced by traditional transposed convolutions~\cite{odena2016deconvolution}.
The choice of asymmetric kernel sizes is due to the fact that a higher dimension along the frequency band axis, may help to capture speech patterns which may mitigate noise and reverberation effects~\cite{ernst2018speech}.The U-Net skip connections are obtained by concatenating pairwise the outputs of layer i), ii), iii) with the outputs of layer viii), vii), vi), respectively. The network size consists of a total of $301.097$ parameters, independently of the size of the input used during training.
\subsection{Training Procedure}
Let us consider a set of microphones $\mathcal{M} = \{\mathbf{m}_m|m=1,\ldots,M\}$ and a set of sources $\mathcal{S} = \{\mathbf{s}_{n_s} |n_s = 1,\ldots,N_s\}$. For each couple of microphones $(\mathbf{m}_i,\mathbf{m}_j) \in \mathcal{M},~s.t.~ i \neq j$ and source $\mathbf{s}_{n_s} \in \mathcal{S}$, we compute the corresponding FS-GCC in anechoic conditions $|{\mathbf{R}}_{ij}|_{n_s}$ and in noisy and reverberant ones $|\tilde{{\mathbf{R}}}_{ij}|_{n_s}$, where index $n_s$ indicates that only source $n_s$ is active when the FS-GCC is computed. We then train the network to minimize the following $\mathrm{MSE}$ loss function
\begin{equation}
    \label{eq:loss-function}
    \mathcal{L} =   \left\|\left({|\hat{\mathbf{R}}}_{ij}|_{n_s} - |{\mathbf{R}}_{ij}|_{n_s}\right)\right\|_2,
\end{equation}
where $|\hat{\mathbf{R}}_{ij}|_{n_s} = \mathcal{U}(|\tilde{\mathbf{R}}_{ij}|_{n_s})$, corresponds to the FS-GCC reconstructed by the network. 
We chose Adam as optimizer, using the default parameters presented in~\cite{kingma2014adam} and we trained the network for a total of $17$ epochs. The validation set was chosen as $20\%$ of the training. To prevent overfitting we applied the Early Stopping regularization technique, which stops the training after $10$ epochs where no improvement on the validation loss is shown. We also reduced the optimizer learning rate by a factor of $1/2$ after $3$ consecutive epochs with no loss improvement. 

%\subsection{Testing and Deployment}
\section{results}
\label{sec:results}

\begin{figure*}[t]
\begin{minipage}[t]{.24\linewidth}
  \centering
  {\pgfplotstableread{figures/results/ANOM_svd_SNR_20.txt}
\SVD%
\pgfplotstableread{figures/results/ANOM_wsvd_SNR_20.txt}
\WSVD%
\pgfplotstableread{figures/results/ANOM_predicted_SNR_20.txt}
\CNN%
\pgfplotstableread{figures/results/ANOM_gcc_SNR_20.txt}
\GCC%

\begin{tikzpicture}
	\begin{axis}[
	    enlargelimits=false,
    	legend columns=1,
        legend style={at={(1,0)},anchor= south east,font=\tiny,mark size = 2pt, rounded corners=1pt,row sep=-0.05cm},
        xmin=0.2,xmax=1,
	  	width=\columnwidth,
        height=\columnwidth,
	  	xlabel={$T60$},
	    %x unit=\second,
	    ylabel={$\mathrm{P}_{\hat{\tau}}$},
	    y unit=\percent,
	    grid=major,
	    legend entries={\tiny{SVD FS-GCC},\tiny{WSVD FS-GCC},\tiny{CNN FS-GCC},\tiny{Conv. GCC}},
	    ]
    \addplot[color=svd_color,mark=diamond*,mark size=2pt,mark options={solid,fill=svd_color}]table[x=0,y=1] from \SVD;
    \addplot[color=wsvd_color,mark=*,mark size=2pt,mark options={solid,fill=wsvd_color}]table[x=0,y=1] from \WSVD;
    \addplot[color=cnn_color,mark=pentagon*,mark size=2pt,mark options={solid,fill=cnn_color}]table[x=0,y=1] from \CNN;
    \addplot[color=gcc_color,mark=square*,mark size=2pt,mark options={solid,fill=gcc_color}]table[x=0,y=1] from \GCC;

    \end{axis}
\end{tikzpicture}}
  %\vspace{1.5cm}
  {(a) Probability of anomalous estimates}\medskip
\end{minipage}
\begin{minipage}[t]{.24\linewidth}
  \centering
    {\pgfplotstableread{figures/results/PK_SNR_svd_SNR_20.txt}
\SVD%
\pgfplotstableread{figures/results/PK_SNR_wsvd_SNR_20.txt}
\WSVD%
\pgfplotstableread{figures/results/PK_SNR_predicted_SNR_20.txt}
\CNN%
\pgfplotstableread{figures/results/PK_SNR_gcc_SNR_20.txt}
\GCC%

\begin{tikzpicture}
	\begin{axis}[
	    enlargelimits=false,
    	legend columns=2,
        legend style={at={(1,1)},anchor=north east,font=\tiny,mark size = 2pt},
        xmin=0.2,xmax=1,
	  	width=\columnwidth,
        height=\columnwidth,
	  	xlabel={$T60$},
	    x unit=\second,
	    ylabel={$\rho_{\hat{\tau}}$},
	    y unit=\decibel,
	    grid=major,
	    ]
 	
    \addplot[color=svd_color,mark=diamond*,mark size=2pt,mark options={solid,fill=svd_color}]table[x=0,y=1] from \SVD;
    \addplot[color=wsvd_color,mark=*,mark size=2pt,mark options={solid,fill=wsvd_color}]table[x=0,y=1] from \WSVD;
    \addplot[color=cnn_color,mark=pentagon*,mark size=2pt,mark options={solid,fill=cnn_color}]table[x=0,y=1] from \CNN;
    \addplot[color=gcc_color,mark=square*,mark size=2pt,mark options={solid,fill=gcc_color}]table[x=0,y=1] from \GCC;

    \end{axis}
\end{tikzpicture}}
 %\vspace{1.5cm}
  {(b) GCC peak SNR \quad(nonanomalous)}\medskip
\end{minipage}
\begin{minipage}[t]{.24\linewidth}
  \centering
  \centerline{\pgfplotstableread{figures/results/MAE_svd_SNR_20.txt}
\SVD%
\pgfplotstableread{figures/results/MAE_wsvd_SNR_20.txt}
\WSVD%
\pgfplotstableread{figures/results/MAE_predicted_SNR_20.txt}
\CNN%
\pgfplotstableread{figures/results/MAE_gcc_SNR_20.txt}
\GCC%

\begin{tikzpicture}
	\begin{axis}[
	    enlargelimits=false,
    	legend columns=2,
        legend style={at={(1,1)},anchor=north east,font=\tiny,mark size = 2pt},
        xmin=0.2,xmax=1,
	  	width=\columnwidth,
        height=\columnwidth,
	  	xlabel={$T60$},
	    x unit=\second,
	    ylabel={$\mathrm{MAE}_{\hat{\tau},\text{na}}~ [\mathrm{samples}]$},
	    %y unit=\metre,
	    grid=major,
	    ]
 	
    \addplot[color=svd_color,mark=diamond*,mark size=2pt,mark options={solid,fill=svd_color}]table[x=0,y=1] from \SVD;
    \addplot[color=wsvd_color,mark=*,mark size=2pt,mark options={solid,fill=wsvd_color}]table[x=0,y=1] from \WSVD;
    \addplot[color=cnn_color,mark=pentagon*,mark size=2pt,mark options={solid,fill=cnn_color}]table[x=0,y=1] from \CNN;
    \addplot[color=gcc_color,mark=square*,mark size=2pt,mark options={solid,fill=gcc_color}]table[x=0,y=1] from \GCC;

    \end{axis}
\end{tikzpicture}}
  %\vspace{1.5cm}
  {(c) Mean absolute error (nonanomalous)}\medskip
\end{minipage}
\begin{minipage}[t]{.24\linewidth}
  \centering
    {\pgfplotstableread{figures/results/SDAE_svd_SNR_20.txt}
\SVD%
\pgfplotstableread{figures/results/SDAE_wsvd_SNR_20.txt}
\WSVD%
\pgfplotstableread{figures/results/SDAE_predicted_SNR_20.txt}
\CNN%
\pgfplotstableread{figures/results/SDAE_gcc_SNR_20.txt}
\GCC%

\begin{tikzpicture}
	\begin{axis}[
	    enlargelimits=false,
    	legend columns=2,
        legend style={at={(1,1)},anchor=north east,font=\tiny,mark size = 2pt},
        xmin=0.2,xmax=1,
	  	width=\columnwidth,
        height=\columnwidth,
	  	xlabel={$T60$},
	    %x unit=\second,
	    ylabel={$\mathrm{SDAE}_{\hat{\tau},\text{na}}[\text{samples}]$},
	    %y unit=\percent,
	    grid=major,
	    ]
 	
    \addplot[color=svd_color,mark=diamond*,mark size=2pt,mark options={solid,fill=svd_color}]table[x=0,y=1] from \SVD;
    \addplot[color=wsvd_color,mark=*,mark size=2pt,mark options={solid,fill=wsvd_color}]table[x=0,y=1] from \WSVD;
    \addplot[color=cnn_color,mark=pentagon*,mark size=2pt,mark options={solid,fill=cnn_color}]table[x=0,y=1] from \CNN;
    \addplot[color=gcc_color,mark=square*,mark size=2pt,mark options={solid,fill=gcc_color}]table[x=0,y=1] from \GCC;

    \end{axis}
\end{tikzpicture}}
 %\vspace{1.5cm}
  {(d) Standard deviation of error (nonanomalous)}\medskip
\end{minipage}

\caption{Performance evaluation with respect to $\mathrm{T60}$ in a $[6~\mathrm{m} \times7~\mathrm{m} \times 3~\mathrm{m}]$ room, $\mathrm{SNR}=20~\mathrm{dB}$. %The fitting curve is a third order polynomial.
}
\label{fig:t60_results_same_room}
\end{figure*}

\begin{figure*}[t]
\begin{minipage}[t]{.24\linewidth}
  \centering
  {\pgfplotstableread{figures/results/ANOM_svd_SNR_20_different_room.txt}
\SVD%
\pgfplotstableread{figures/results/ANOM_wsvd_SNR_20_different_room.txt}
\WSVD%
\pgfplotstableread{figures/results/ANOM_predicted_SNR_20_different_room.txt}
\CNN%
\pgfplotstableread{figures/results/ANOM_gcc_SNR_20_different_room.txt}
\GCC%

\begin{tikzpicture}
	\begin{axis}[
	    enlargelimits=false,
    	legend columns=1,
        legend style={at={(1,0)},anchor= south east,font=\tiny,mark size = 2pt, rounded corners=1pt, row sep=-0.05cm},
        xmin=0.2,xmax=1,
	  	width=\columnwidth,
        height=\columnwidth,
	  	xlabel={$T60$},
	    %x unit=\second,
	    ylabel={$\mathrm{P}_{\hat{\tau}}$},
	    y unit=\percent,
	    grid=major,
	    legend entries={\tiny{SVD FS-GCC},\tiny{WSVD FS-GCC},\tiny{CNN FS-GCC},\tiny{Conv. GCC}},
	    ]
 	
    \addplot[color=svd_color,mark=diamond*,mark size=2pt,mark options={solid,fill=svd_color}]table[x=0,y=1] from \SVD;
    \addplot[color=wsvd_color,mark=*,mark size=2pt,mark options={solid,fill=wsvd_color}]table[x=0,y=1] from \WSVD;
    \addplot[color=cnn_color,mark=pentagon*,mark size=2pt,mark options={solid,fill=cnn_color}]table[x=0,y=1] from \CNN;
    \addplot[color=gcc_color,mark=square*,mark size=2pt,mark options={solid,fill=gcc_color}]table[x=0,y=1] from \GCC;

    \end{axis}
\end{tikzpicture}}
  %\vspace{1.5cm}
  {(a) Probability of anomalous estimates}\medskip
\end{minipage}
\begin{minipage}[t]{.24\linewidth}
  \centering
    {\pgfplotstableread{figures/results/PK_SNR_svd_SNR_20_different_room.txt}
\SVD%
\pgfplotstableread{figures/results/PK_SNR_wsvd_SNR_20_different_room.txt}
\WSVD%
\pgfplotstableread{figures/results/PK_SNR_predicted_SNR_20_different_room.txt}
\CNN%
\pgfplotstableread{figures/results/PK_SNR_gcc_SNR_20_different_room.txt}
\GCC%

\begin{tikzpicture}
	\begin{axis}[
	    enlargelimits=false,
    	legend columns=2,
        legend style={at={(1,1)},anchor=north east,font=\tiny,mark size = 2pt},
        xmin=0.2,xmax=1,
	  	width=\columnwidth,
        height=\columnwidth,
	  	xlabel={$T60$},
	    x unit=\second,
	    ylabel={$\rho_{\hat{\tau}}$},
	    y unit=\decibel,
	    grid=major,
	    ]
 	
    \addplot[color=svd_color,mark=diamond*,mark size=2pt,mark options={solid,fill=svd_color}]table[x=0,y=1] from \SVD;
    \addplot[color=wsvd_color,mark=*,mark size=2pt,mark options={solid,fill=wsvd_color}]table[x=0,y=1] from \WSVD;
    \addplot[color=cnn_color,mark=pentagon*,mark size=2pt,mark options={solid,fill=cnn_color}]table[x=0,y=1] from \CNN;
    \addplot[color=gcc_color,mark=square*,mark size=2pt,mark options={solid,fill=gcc_color}]table[x=0,y=1] from \GCC;

    \end{axis}
\end{tikzpicture}}
 %\vspace{1.5cm}
  {(b) GCC peak SNR \quad(nonanomalous)}\medskip
\end{minipage}
\begin{minipage}[t]{.24\linewidth}
  \centering
  \centerline{\pgfplotstableread{figures/results/MAE_svd_SNR_20_different_room.txt}
\SVD%
\pgfplotstableread{figures/results/MAE_wsvd_SNR_20_different_room.txt}
\WSVD%
\pgfplotstableread{figures/results/MAE_predicted_SNR_20_different_room.txt}
\CNN%
\pgfplotstableread{figures/results/MAE_gcc_SNR_20_different_room.txt}
\GCC%

\begin{tikzpicture}
	\begin{axis}[
	    enlargelimits=false,
    	legend columns=2,
        legend style={at={(1,1)},anchor=north east,font=\tiny,mark size = 2pt},
        xmin=0.2,xmax=1,
	  	width=\columnwidth,
        height=\columnwidth,
	  	xlabel={$T60$},
	    x unit=\second,
	    ylabel={$\mathrm{MAE}_{\hat{\tau},\text{na}}~ [\mathrm{samples}]$},
	    %y unit=\metre,
	    grid=major,
	    ]
 	
    \addplot[color=svd_color,mark=diamond*,mark size=2pt,mark options={solid,fill=svd_color}]table[x=0,y=1] from \SVD;
    \addplot[color=wsvd_color,mark=*,mark size=2pt,mark options={solid,fill=wsvd_color}]table[x=0,y=1] from \WSVD;
    \addplot[color=cnn_color,mark=pentagon*,mark size=2pt,mark options={solid,fill=cnn_color}]table[x=0,y=1] from \CNN;
    \addplot[color=gcc_color,mark=square*,mark size=2pt,mark options={solid,fill=gcc_color}]table[x=0,y=1] from \GCC;

    \end{axis}
\end{tikzpicture}}
  %\vspace{1.5cm}
  {(c) Mean absolute error (nonanomalous)}\medskip
\end{minipage}
\begin{minipage}[t]{.24\linewidth}
  \centering
    {\pgfplotstableread{figures/results/SDAE_svd_SNR_20_different_room.txt}
\SVD%
\pgfplotstableread{figures/results/SDAE_wsvd_SNR_20_different_room.txt}
\WSVD%
\pgfplotstableread{figures/results/SDAE_predicted_SNR_20_different_room.txt}
\CNN%
\pgfplotstableread{figures/results/SDAE_gcc_SNR_20_different_room.txt}
\GCC%

\begin{tikzpicture}
	\begin{axis}[
	    enlargelimits=false,
    	legend columns=2,
        legend style={at={(1,1)},anchor=north east,font=\tiny,mark size = 2pt},
        xmin=0.2,xmax=1,
	  	width=\columnwidth,
        height=\columnwidth,
	  	xlabel={$T60$},
	    %x unit=\second,
	    ylabel={$\mathrm{SDAE}_{\hat{\tau},\text{na}}[\text{samples}]$},
	    %y unit=\percent,
	    grid=major,
	    ]
 	
    \addplot[color=svd_color,mark=diamond*,mark size=2pt,mark options={solid,fill=svd_color}]table[x=0,y=1] from \SVD;
    \addplot[color=wsvd_color,mark=*,mark size=2pt,mark options={solid,fill=wsvd_color}]table[x=0,y=1] from \WSVD;
    \addplot[color=cnn_color,mark=pentagon*,mark size=2pt,mark options={solid,fill=cnn_color}]table[x=0,y=1] from \CNN;
    \addplot[color=gcc_color,mark=square*,mark size=2pt,mark options={solid,fill=gcc_color}]table[x=0,y=1] from \GCC;

    \end{axis}
\end{tikzpicture}}
 %\vspace{1.5cm}
  {(d) Standard deviation of error (nonanomalous)}\medskip
\end{minipage}

\caption{Performance evaluation with respect to $\mathrm{T60}$ in a $[9~\mathrm{m}\times8~\mathrm{m}\times4~\mathrm{m}]$ room, with $\mathrm{SNR}=20~\mathrm{dB}$. %The fitting curve is a third order polynomial.
}
\label{fig:t60_results_different_room}
\end{figure*}

In this section we present simulation results demonstrating the effectiveness of the described method. In particular we show how TDoA estimation results vary w.r.t the reverberation time, testing both in the same room used for training as well as in a different one.
\subsection{Evaluation Measures}
A TDoA estimate is classified as an anomaly according to its absolute error $e = |\tau -\hat{\tau}|$, where $\tau$ is the true TDoA, while $\hat{\tau}$ is the estimate. If $e_i> T_c/2$ the estimate is considered anomalous, where $T_c$ is the signal correlation time~\cite{champagne1996performance}. In the case of the speech signals we used for testing, $ 12< T_c < 30$. The TDoA estimation performances were evaluated in percentage of anomalous estimates over total estimates ($P_{\hat{\tau}}$) and GCC peak $\mathrm{SNR}$ ($\rho_{\hat{\tau},\text{na}}$), mean absolute error ($\mathrm{MAE}$) and standard deviation ($\mathrm{SDAE}_{\hat{\tau},\text{na}}$), for the subset of nonanomalous estimates \cite{Cobos2019FSGCC}. 

\subsection{Simulation Setup}
To generate the training data we used 80 two-microphone arrays and 80 random source positions on the plane ($x,y,z=1.25~\mathrm{m}$). The SNR was randomly chosen between $-10~\mathrm{dB}$ and $20~\mathrm{dB}$ while the $\mathrm{T60}$ varied between $0.2~\mathrm{s}$ and $1~\mathrm{s}$. The test set consisted of $5$ two-microphone arrays and $5$ sources positions, for a total of $25$ setups. 
The speech signals used for training and testing were drawn from LibriSpeech~\cite{panayotov2015librispeech} and they are all 2 $\mathrm{s} $ long. We used $1$ speaker for training and $9$ for testing.% making sure that each source was able to emit the signal related to all the speakers. 
The room considered for training was of size $[6 ~\mathrm{m}\times 7 ~\mathrm{m}\times 3~\mathrm{m}]$. During testing we considered both the same room used for training as well as a different one of size  $[9~\mathrm{m}\times 8~\mathrm{m} \times 4~\mathrm{m}]$, while keeping the same relative positions for sources and microphone arrays. The synthetic microphone signals were obtained by convolving the source signals with the impulse response generated through the image-source method~\cite{habets2006room}. Then, the signals were processed by the different methods in the STFT domain, using a sampling frequency $f_s=44100~\mathrm{Hz}$, a frame length of 2048 samples and Hann windowing with $75\%$ overlap. In the training set we considered just one frame for each source, for a total size of $N^{\text{train}}=80\times80=6400$ examples. In the test set, instead, for each source we considered all $177$ frames. The total number of estimates used to evaluate each method at each $\mathrm{T60}$ value is of $N^{\text{test}}=5\time 5 \times 177 = 4425$ estimates.
The FS-GCC parameters were $B = 128$, $M=29$, resulting in $L=32$ frequency bands.

\subsection{TDoA estimation results}
In Fig.~\ref{fig:t60_results_same_room} we show results regarding TDoA estimation using the test set computed in the same room used for training.  As it is clear from Fig~\ref{fig:t60_results_same_room}(a), the method presented in this paper (CNN FS-GCC), is the one that produces the lowest number of anomalous estimates for all $\mathrm{T60}$ values. WSVD FS-GCC obtains slightly worse performance, while both SVD FS-GCC and GCC produce a sensibly higher number of anomalies. As shown in Fig.~\ref{fig:t60_results_same_room}(b), CNN FS-GCC is the method that shows the highest GCC peak SNR performance, while GCC performs worst. GCC has the best performances in terms of both $\mathrm{SDAE}_{\hat{\tau},\text{na}}$ (Fig.~\ref{fig:t60_results_same_room}(c)) and $\mathrm{MAE}_{\hat{\tau},\text{na}}$(Fig~\ref{fig:t60_results_same_room}(d)). Note, however, that these measures are only computed for nonanomalous estimates, which occur less frequently for GCC than for all the other methods and, thus, are less informative in this case.
%which however are relative only to nonanomalous estimates,
%Fig.~\ref{fig:t60_results_same_room}(a) these number less for GCC w.r.t. all the other considered techniques, thus making $\mathrm{MAE}$ and $\mathrm{SDAE}$ results less informative.
In order to show the ability of our network to generalize, in Fig.~\ref{fig:t60_results_different_room}, we show TDoA results relative to a room different from the training one, the curves follow the same patterns of the ones shown in Fig.~\ref{fig:t60_results_same_room}. In both cases, we repeated the tests also for $\mathrm{SNR}$ values $10~\mathrm{dB}$,~$0~\mathrm{dB}$ and $-10~\mathrm{dB}$. The patterns follow the same curves as the ones shown in Fig.~\ref{fig:t60_results_same_room} and Fig.~\ref{fig:t60_results_different_room}, obtaining gradually worse performance with lower $\mathrm{SNR}$ values. Finally, although not evaluated in this paper, the higher performance of CNN FS-GCC in terms of Peak SNR may contribute to better localization results using Steered Response Power (SRP) approaches \cite{CobosSPL2011}, as confirmed in \cite{Cobos2019FSGCC} for the other FS-GCC methods.

\section{Conclusions}
\label{sec:conclusions}
In this paper we have shown an approach for TDoA estimation, based on a U-Net fully convolutional denoising autoencoder applied to the FS-GCC, a recently developed frequency-sliding version of the GCC.
The goal is to reconstruct the FS-GCC corresponding to an anechoic environment from a FS-GCC affected by both  noise and reverberation. Through a simulation campaign we showed the effectiveness of our method compared to GCC and FS-GCC approaches where no CNNs are applied. We also demonstrated the ability of our network to generalize, by training in a room and testing in a different one.
The results presented in this work stimulate us to a series of future developments, such as the use of different network architectures like Generative Adversarial Networks and the extension of the generalization capabilities in order to several microphone spacings and multi-source scenarios. The improvement of this method could lead to robust acoustic source localization methods.

% Below is an example of how to insert images. Delete the ``\vspace'' line,
% uncomment the preceding line ``\centerline...'' and replace ``imageX.ps''
% with a suitable PostScript file name.
% -------------------------------------------------------------------------
%\begin{figure}[htb]

%
%\begin{minipage}[b]{1.0\linewidth}
%  \centering
%  \centerline{\includegraphics[width=8.5cm]{image1}}
%  \vspace{2.0cm}
%  \centerline{(a) Result 1}\medskip
%\end{minipage}
%
%\begin{minipage}[b]{.48\linewidth}
%  \centering
%  \centerline{\includegraphics[width=4.0cm]{image3}}
%  \vspace{1.5cm}
%  \centerline{(b) Results 3}\medskip
%\end{minipage}
%\hfill
%\begin{minipage}[b]{0.48\linewidth}
%  \centering
%  \centerline{\includegraphics[width=4.0cm]{image4}}
%  \vspace{1.5cm}
%  \centerline{(c) Result 4}\medskip
%\end{minipage}
%
%\caption{Example of placing a figure with experimental results.}
%\label{fig:res}
%
%\end{figure}

%\vfill
%\pagebreak

\vfill\pagebreak

\bibliographystyle{IEEEbib}
\balance
\bibliography{strings,refs}

\end{document}